\documentclass{JHEP3} 

\usepackage{epsfig}
\usepackage{amsmath, amssymb}
\usepackage{concmath, palatino}
\usepackage{mathrsfs}
\usepackage{epic}
\usepackage{color}

\newcommand\fverb{\setbox\pippobox=\hbox\bgroup\verb}
\newcommand\fverbdo{\egroup\medskip\noindent%
			\fbox{\unhbox\pippobox}\ }
\newcommand\fverbit{\egroup\item[\fbox{\unhbox\pippobox}]}
\newbox\pippobox

\newcommand{\be}{\begin{equation}}
\newcommand{\ee}{\end{equation}}
\newcommand{\ba}{\begin{eqnarray}}
\newcommand{\ea}{\end{eqnarray}}

\newcommand{\ads}{AdS_5\times S^5}

\newcommand{\ddb}{{\overline{\mathscr D}}}
\newcommand{\sym}{$\mathcal{N}=4$ SYM }

\allowdisplaybreaks

    \newcommand{\beq}{\begin{equation}}
    \newcommand{\eeq}{\end{equation}}
    \newcommand\beqa{\begin{eqnarray}}
    \newcommand\eeqa{\end{eqnarray}}

\title{Large spin expansion of the wrapping correction to
          Freyhult-Rej-Zieme twist operators}

\author{Matteo Beccaria\\
  Dipartimento di Fisica, Universita' del Salento, 
  Via Arnesano, 73100 Lecce \&\\
  INFN, Sezione di Lecce\\
  E-mail: \email{matteo.beccaria$\bullet$le.infn.it}
}

\author{Guido Macorini\\
  Dipartimento di Fisica, Universita' del Salento, 
  Via Arnesano, 73100 Lecce \&\\
  INFN, Sezione di Lecce\\
  E-mail: \email{guido.macorini$\bullet$le.infn.it}
}

\author{CarloAlberto Ratti\\
  Dipartimento di Fisica, Universita' del Salento, 
  Via Arnesano, 73100 Lecce \&\\
  INFN, Sezione di Lecce\\
  E-mail: \email{carloalberto.ratti$\bullet$le.infn.it}
}

\abstract{ Twist operators in the closed $\mathfrak{sl}(2)$ sector of
  planar $\mathcal N=4$ SYM are characterized by their spin. The
  explicit dependence of anomalous dimensions on this important
  parameter is a source of interesting information.  Wrapping
  corrections are a non trivial part of the calculation and are under
  control in the framework of thermodynamical Bethe Ansatz valid for
  the full theory and thoroughly checked in that sector. The extension
  to more general twist operators beyond $\mathfrak{sl}(2)$ has been
  recently accomplished for the so-called 3-gluon operators that are a
  special case of the generalized twist operators introduced by Freyhult, Rej and Zieme.
 Such operators are dual to spinning strings
  configurations with two spins $S_{1}$, $S_{2}$ in $AdS_5$ and charge
  in $S^5$.  We compute the expansion of the weak-coupling leading
  order wrapping correction in the gauge theory limit dual to large
  $S_{1}$ and fixed $S_{2}$. We present a simple algorithm for the
  calculation and provide explicit results illustrating the general
  structure of the expansion.  }

\begin{document} 

\section{Introduction}

The computation of finite size corrections to states/operators in
AdS/CFT correspondence is an important technical issue. Recently, in
the integrable planar limit, this problem has been solved in full
generality by means of the mirror thermodynamic Bethe Ansatz developed
for the $\ads$ superstring in~\cite{TBA1}. Formerly, the associated
Y-system had been proposed in \cite{GKV} based on symmetry arguments
and educated guesses about the analyticity and asymptotic properties
of the Y-functions. The predicted finite size corrections have been
deeply tested in \cite{TBA2}, mainly in the closed $\mathfrak{sl}(2)$
subsector.  The relevant operators are represented by the insertion of
$n$ covariant derivatives $\mathscr{D}$ into the protected half-BPS
state $\mbox{Tr} \mathcal{Z}^{L}$ ($\mathcal Z$ being one of the three
complex scalars of $\mathcal N=4$ SYM)
\begin{equation}
  \mathbb{O}_{n, L}^{\mathcal Z} = \sum_{s_{1},\dots, s_{L}} c_{s_{1}, \dots, s_{L}}\,
  \mbox{Tr} \left( \mathscr{D}^{s_1}\mathcal{Z}\cdots\mathscr{D}^{s_L}\mathcal{Z}\right),
  \quad  \mathrm{with} \quad n=s_1+\cdots+s_{L}\,. 
\end{equation}
Their anomalous dimensions can be obtained from a non-compact,
length-$L$ $\mathfrak{sl}(2)$ invariant integrable spin chain with $n$
excitations. The interaction range between scattering magnons
increases with the perturbative order.  As soon as it exceeds the
length of the spin chain and {\it wraps} around it, the S-matrix
picture fails and no asymptotic region can be defined any longer.  For
length $L$ operators this effect is delayed by superconformal
invariance and starts at order $g^{2L+4}$. In this regime, wrapping
corrections cannot be obtained within the asymptotic Bethe Ansatz and
require the full use of the thermodynamical Bethe Ansatz framework.
The most accurate available calculations are at five-loop order for
the special length $L=2$~\cite{Lukowski:2009ce} and at six-loop order
for $L=3$~\cite{Velizhanin:2010cm}. In these cases, the minimal
anomalous dimension of $\mathbb{O}_{n, L}^{\mathcal Z}$ can be
obtained in closed form as a function of the number of excitations
$n$.

\medskip The availability of $n$ as a control parameter is a
remarkable fact since it opens the door to very interesting cross
checks of the calculations. For instance, at large $n$, it is found
that a generalized Gribov-Lipatov reciprocity (see \cite{reciprocity}
and the recent review \cite{Beccaria:2010tb}) holds predicting half of
the large $n$ expansion in terms of the other half.  Also, in the
twist-2 case, the analytic continuation in the spin parameter $n$
allows to test the predictions of the BFKL
equation~\cite{Lipatov:1976zz} governing the poles around unphysical
negative values of $n$.

\medskip Apart from these important tests, the computation of wrapping
corrections as functions of (or series expansions in) the parameter
$n$ is also very useful in order to predict general features. For
instance, a recurring theme in AdS/CFT is the assumption that wrapping
corrections are somewhat suppressed at large $n$~\footnote{In all
  known cases they scale like $\frac{1}{n^{2}}$ with possible
  enhancement factors growing like powers of $\log n$.}. This permits,
in first approximation, to neglect them. A nice example where such an
approximation is needed is the computation of Maldacena {\em et al.}
of the two loop expressions for polygonal Wilson loops expectation
values \cite{Gaiotto:2010fk} . It is based on an operator product
expansion where the spectrum of excitations of the flux tube
stretching between two null Wilson lines can also be viewed as the
spectrum of excitations around the infinite spin limit of finite twist
operators in the $\mathfrak{sl}(2)$ sector of \sym, or the GKP
\cite{Gubser:2002tv} string.  Integrability and AdS/CFT correspondence
effectively help in computing such spectrum and wrapping corrections
are assumed to be negligible. Such a statement is safe for the GKP
background, but is only a conjecture (although reasonable) for the
excitations over the GKP string.

\medskip Thus, generally speaking, it is clearly important to increase
our knowledge of the structure of wrapping corrections to twist
operators beyond the simple $\mathfrak{sl}(2)$ sector~\footnote{ Here,
  the loose term {\em twist operator} refers to gauge invariant
  composite operators built with a fixed number of elementary fields
  and an increasing number of covariant derivatives acting on
  them. }. Such an extension has been recently presented in
\cite{BMR11} where we studied wrapping corrections to operators whose
multi-loop asymptotic contributions had been computed in
~\cite{Beccaria:2007pb}. We shall refer to these operators as 3-gluon
operators~\footnote{At one-loop they have the same form as
  $\mathfrak{sl}(2)$ operators, with the scalar $\mathcal Z$ being
  replaced by a physical gauge field component.}. Indeed, in
\cite{BMR11} we computed the leading order wrapping correction to the
lowest anomalous dimension of such operators in closed form as a
function of $n$.

\medskip Actually, 3-gluon operators are related by superconformal
invariance to a special case of a larger family studied in
\cite{Freyhult:2009fc} which we shall dub Freyhult-Rej-Zieme (FRZ) twist
operators. For the length 3 case we are interested in, they take the
following (schematic) form
\begin{equation}
  \label{operator}
  \mathbb{O}^{\rm FRZ}_{n,m} = \mbox{Tr}\,(\mathscr{D}^{n+m} \Bar{\mathscr{D}}^m \mathcal{Z}^3)+\cdots,
\end{equation}
where dots denote a linear combination of similar terms with the
covariant derivatives spread over the scalar fields.  These operators
reduce to length 3 operators in the $\mathfrak{sl}(2)$ subsector for
$m=0$. For $m=1$ we get descendants of twist--2 operators. For $m=2$
we get the 3-gluon operators.  At strong coupling, the FRZ operators
are duals to minimal energy spinning strings configurations with two
spins $S_1$ and $S_2$ in $AdS_5$ and charge $J$ in $S^5$ given by
\begin{equation}
  S_1 = n+m-\frac{1}{2}~, \qquad S_2 = m-\frac{1}{2}~, \qquad J=L=3.
\end{equation}

\medskip The main result of \cite{Freyhult:2009fc} is the large $n$
expansion of the asymptotic minimal anomalous dimension of
$\mathbb{O}^{\rm FRZ}_{n,m}$ for fixed ratio $n/m$ or fixed $m$.  The
expansion is obtained at all orders in the coupling and including the
leading term $\sim \log n$ as well as the subleading asymptotically
constant correction $\sim n^{0}$. These two contributions are expected
to be free of wrapping corrections.  In this paper, we consider
precisely the leading order wrapping correction which appears at four
loops.  We provide an algorithm to compute its large $n$ expansion for
fixed $m$ and present explicit results for $m=2,3,4$. As we mentioned,
for $m=2$ we have to match the 3-gluon result obtained in
\cite{BMR11}. The expansions for the other two values are new. In full
generality, we prove the $\frac{\log n}{n^{2}}$ scaling behaviour at
large $n$ thus confirming the assumption in
\cite{Freyhult:2009fc}. Notice that for the considered states with
$m>2$ no asymptotic closed form of the anomalous dimension is known
beyond one-loop.

\medskip The plan of the paper is the following. In
Sec.~(\ref{sec:oneloop}), we summarize the one-loop solution of the
Bethe Ansatz equations for FRZ operators. Sec.~(\ref{sec:Y-system})
presents the necessary Y-system formulae for the efficient computation
of the leading order wrapping correction. The algorithm for the
derivation of its large spin expansion is described and tested in
Sec.~(\ref{sec:algorithm}). Our results are summarized in
Sec.~(\ref{sec:summary}) .
  
\section{One loop solution of the FRZ operators}
\label{sec:oneloop}

In this section, we review the one-loop solution of the FRZ states and
give explicit information on the Baxter polynomials entering the
wrapping calculation. In particular, we shall provide the explicit
form of the Baxter polynomials whose degree is independent on the spin
$n$.
  
The excitation pattern of the operators in Eq.~(\ref{operator}) has
the following form in the higher Dynkin diagram of
$\mathfrak{psu}(2,2|4)$ in the $\mathfrak{su}(2)$ grading
\begin{equation}
  \label{eq:dynkin}
  \begin{minipage}{260pt}
    \setlength{\unitlength}{1pt} \small\thicklines
    \begin{picture}(260,55)(-10,-30)
      \put(-72,0){\line(1,0){34}}  
      \put(-30,00){\circle{15}}
      \put(-35,-5){\line(1, 1){10}}  
      \put(-35, 5){\line(1,-1){10}}  
      \dottedline{3}(-22,0)(12,0)    
      \put( 20,00){\circle{15}}     
      \dottedline{3}(60,0)(30,0)   
      \put( 70,00){\circle{15}}
      \put( 65,-5){\line(1, 1){10}}  
      \put( 65, 5){\line(1,-1){10}}  
      \put( 70,-15){\makebox(0,0)[t]{$n+2m-1$}}  
      \put( 77,00){\line(1,0){36}} 
      \put(120,00){\circle{15}}
      \put(120,15){\makebox(0,0)[b]{$+1$}} 
      \put(120,-15){\makebox(0,0)[t]{$n+2m$}} 
      \put(127,00){\line(1,0){36}} 
      \put(170,00){\circle{15}}
      \put(165,-5){\line(1, 1){10}}  
      \put(165, 5){\line(1,-1){10}}  
      \put(170,-15){\makebox(0,0)[t]{$n+2m-2$}} 
      \dottedline{3}(178,0)(212,0) 
      \put(220,00){\circle{15}}
      \put(220,-15){\makebox(0,0)[t]{$m-1$}} 
      \dottedline{3}(228,0)(262,0) 
      \put(270,00){\circle{15}}
      \put(265,-5){\line(1, 1){10}} 
      \put(265, 5){\line(1,-1){10}} 
      \put(278,0){\line(1,0){30}} 
    \end{picture}
  \end{minipage}
\end{equation}
By dualizing the diagram on  node 1, 3 we arrive at the simpler
configuration 
\begin{equation}
  \label{eq:dynkin2}
  \begin{minipage}{260pt}
    \setlength{\unitlength}{1pt} \small\thicklines
    \begin{picture}(260,55)(-10,-30)
      \dottedline{3}(-72,0)(-40,0)  
      \put(-30,00){\circle{15}}
      \put(-35,-5){\line(1, 1){10}}  
      \put(-35, 5){\line(1,-1){10}}  
      \put(-23,00){\line(1,0){35}} 
      \put( 20,00){\circle{15}}     
      \put( 27,00){\line(1,0){35}} 
      \put( 70,00){\circle{15}}
      \put( 65,-5){\line(1, 1){10}}  
      \put( 65, 5){\line(1,-1){10}}  
      \dottedline{3}(78,0)(112,0) 
      \put(120,00){\circle{15}}
      \put(115,-5){\line(1, 1){10}} 
      \put(115, 5){\line(1,-1){10}} 
      \put(120,15){\makebox(0,0)[b]{$+1$}} 
      \put(120,-15){\makebox(0,0)[t]{$n+2m$}} 
      \put(127,00){\line(1,0){34}} 
      \put(170,00){\circle{15}}
      \put(165,-5){\line(1, 1){10}}  
      \put(165, 5){\line(1,-1){10}}  
      \put(170,-15){\makebox(0,0)[t]{$n+2m-2$}} 
      \dottedline{3}(178,0)(212,0) 
      \put(220,00){\circle{15}}
      \put(220,-15){\makebox(0,0)[t]{$m-1$}} 
      \dottedline{3}(228,0)(262,0) 
      \put(270,00){\circle{15}}
      \put(265,-5){\line(1, 1){10}} 
      \put(265, 5){\line(1,-1){10}} 
      \put(278,0){\line(1,0){30}} 
    \end{picture}
  \end{minipage}
\end{equation}

\noindent
The  Bethe equations  in this grading
(\ref{eq:dynkin2}) are~\footnote{
 Shifted quantities are defined as
 \begin{equation}
   F^{\underbrace{\pm\dots\pm}_{a}}(u) = F^{[\pm a]}(u) = F\left(u\pm i\,\frac{a}{2}\right).
 \end{equation}
 $Q_{\ell}$'s are the Baxter polynomials vanishing on the $\ell$-th node roots
 \begin{equation}
   \label{QsBaxter}
   Q_{\ell}(u) = \prod_{i=1}^{K_{\ell}}\left(u-u_{\ell,i}\right),
 \end{equation}
 $K_{\ell}$ being the number of excitations on the $\ell$-th node.
}
\begin{equation}
  \label{Bethe-eq}
  \left( \frac{u^+_{4,k}}{u^-_{4,k}} \right)^3  =  \frac{Q_5^-}{Q_5^+}\bigg|_{u_{4,k}},
  \qquad 1 = \frac{Q_6^+}{Q_6^-}\bigg|_{u_{5,k}} \frac{Q_4^-}{Q_4^+}\bigg|_{u_{5,k}},
  \qquad  - 1  =  \frac{Q_6^{++}}{Q_6^{--}}\bigg|_{u_{6,k}} \frac{Q_5^-}{Q_5^+}\bigg|_{u_{6,k}}.
\end{equation}
The solution to the system (\ref{Bethe-eq}) is explicitly written out
in Appendix~\ref{app:oneloop-solution} where we provide the
expressions of the polynomials $Q_{4,5,6}$. Since the degree of
$Q_{4,5}$ is dependent on $n$, the computation of the large $n$ limit
is definitely non - trivial.

Although the definition of $Q_{6}$ is rather complicated, it is a
polynomial of order $m-1$ whose coefficients depend on $n$. We can
reconstruct them explicitly for general values of $n$ at least for the
first values of $m$. We do this for $m=2,3,\dots,8$. The polynomials
are
\begin{eqnarray}
  \label{FirstQ6}
  &&
  Q_{6}^{m=2}= u,
  \nonumber \\
  &&
  Q_6^{m=3}=  4 (n+3)(n+8) u^2 +32 + 11 n + n^2,
  \nonumber \\
  &&
  Q_6^{m=4}=  \left(132+4 n (n+14)\right)u^3 +\left(213 + 4 n (n+14)\right) u,
  \nonumber \\
  &&
  Q_6^{m=5}= 
  16 (n+3)(n+5)(n+12)(n+14)u^4
  \nonumber \\
  &&
  \qquad \qquad 
  +8 (n+5) (n+12)(402+5n(n+17)) u^2 
  \nonumber \\
  &&
  \qquad \qquad 
  + 47256 + 3n (n+17) (434 + 3 n (n+17)),
  \nonumber \\
  &&
  Q_6^{m=6}= 
  16 (n+3)(n+5)(n+15)(n+17)u^5
  \nonumber \\
  &&
  \qquad \qquad 
  + (681000+40n(n+20)(377+2n(n+20)))u^3 
  \nonumber \\
  &&
  \qquad \qquad 
  + (670425+8n(n+20)(1633+8n(n+20))) u ,
  \nonumber \\
  &&
  Q_6^{m=7}= 
  64 (n+3) (n+5) (n+7) (n+16) (n+18) (n+20) u^6
   \\
  &&
  \qquad \qquad 
  +80 (n+5) (n+7) (n+16) (n+18) (7 n (n+23)+1068) u^4
  \nonumber \\
  &&
  \qquad \qquad 
  +4 (n+7) (n+16) (n (n+23) (259 n (n+23)+71250)+4936680) u^2
  \nonumber \\
  &&
  \qquad \qquad 
  +45 (n (n+23) (n (n+23) (5 n (n+23)+1958)+255720)+11140992),
  \nonumber \\
  && 
  Q_6^{m=8}= 64 (n+3) (n+5) (n+7) (n+19) (n+21) (n+23) u^7
  \nonumber \\
  && \qquad \qquad 
  +112 (n+5) (n+7) (n+19) (n+21) (8 n (n+26)+1581) u^5
  \nonumber \\
  && \qquad \qquad 
  +196 (n+7) (n+19) (4 n (n+26) (4 n (n+26)+1431)+516495) u^3
  \nonumber \\
  && \qquad \qquad 
  +3 (4 n (n+26) (16 n (n+26) (12 n (n+26)+6085)+16477937)+3724800415) u.
  \nonumber 
\end{eqnarray}
These results will be useful in the following since they are explicit
in $n$ and can be used to extract large $n$ contributions.

\medskip In order to efficiently evaluate wrapping corrections it is
convenient to dualize the diagram (\ref{eq:dynkin2}) at nodes 5 and
7. We get a configuration where the number of roots at nodes 5, 6, 7
does not depend anymore on $n$. More precisely, we get
\begin{eqnarray}
  \label{eq:dynkin3}
  \begin{minipage}{260pt}
    \setlength{\unitlength}{1pt} \small\thicklines
    \begin{picture}(260,55)(-10,-30)
      \dottedline{3}(-72,0)(-40,0)  
      \put(-30,00){\circle{15}}
      \put(-35,-5){\line(1, 1){10}}  
      \put(-35, 5){\line(1,-1){10}}  
      \put(-23,00){\line(1,0){35}} 
      \put( 20,00){\circle{15}}     
      \put( 27,00){\line(1,0){35}} 
      \put( 70,00){\circle{15}}
      \put( 65,-5){\line(1, 1){10}}  
      \put( 65, 5){\line(1,-1){10}}  
      \dottedline{3}(78,0)(112,0) 
      \put(120,00){\circle{15}}
      \put(120,15){\makebox(0,0)[b]{$+1$}} 
      \put(120,-15){\makebox(0,0)[t]{$n+2m$}} 
      \dottedline{3}(128,0)(162,0) 
      \put(170,00){\circle{15}}
      \put(165,-5){\line(1, 1){10}}  
      \put(165, 5){\line(1,-1){10}}  
      \put(170,-15){\makebox(0,0)[t]{$m$}} 
      \put(177,00){\line(1,0){35}} 
      \put(220,00){\circle{15}}
      \put(220,-15){\makebox(0,0)[t]{$m-1$}} 
      \put(227,00){\line(1,0){35}} 
      \put(270,00){\circle{15}}
      \put(265,-5){\line(1, 1){10}} 
      \put(265, 5){\line(1,-1){10}} 
      \put(270,-15){\makebox(0,0)[t]{$m-2$}} 
      \dottedline{3}(310,0)(280,0) 
    \end{picture}
  \end{minipage}
\nonumber \\
\end{eqnarray}
This is the direct extension of eq. (3.13) in \cite{BMR11}. An
important difference are the $m-2$ roots appearing on node 7.

The one--loop Bethe equations are now
\begin{equation}
  \label{Bethe-eq-sl(2)}
  -\left( \frac{u^+_{4,k}}{u^-_{4,k}} \right)^3 = \frac{Q_4^{--}}{Q_4^{++}}\bigg|_{u_{4,k}} \frac{\widetilde{Q}_5^+}{\widetilde{Q}_5^-}\bigg|_{u_{4,k}},\quad
  1= \frac{Q_6^+}{Q_6^-}\bigg|_{u_{5,k}} \frac{Q_4^-}{Q_4^+}\bigg|_{\widetilde{u}_{5,k}},
  \quad
  1 =  \frac{\widetilde{Q}_5^-}{\widetilde{Q}_5^+}\bigg|_{u_{6,k}},
  \quad
  1 =  \frac{Q_6^-}{Q_6^+}\bigg|_{\widetilde{u}_{7,k}},
\end{equation}
where the dual Baxter functions $\widetilde{Q}_5$, $\widetilde{Q}_7$
are defined by
\begin{eqnarray}
  \label{dualBaxter}
  &&
  \widetilde{Q}_5 Q_5 = Q_4^+ Q_6^- -Q_4^- Q_6^+, \qquad \qquad
  \widetilde{Q}_7 = Q_6^+ - Q_6^-.
\end{eqnarray}
They are polynomials of order $m$ and $m-2$ respectively.  Explicitly,
for $m=2,3,\dots,8$ they read
\begin{eqnarray}
  \label{FirstQ5}
  &&
  \widetilde{Q}_5^{m=2}=  (n+3)(n+5)u^2 + \frac{1}{4}(n+4)^2,
  \nonumber \\
  &&
  \widetilde{Q}_5^{m=3}= \frac{1}{2} \left(
    (n+3)(n+8) u^3+(n+5)(n+6) u \right),
  \nonumber \\
  &&
  \widetilde{Q}_5^{m=4}= \frac{1}{16} \left( 
    +16(n+3)(n+5)(n+9)(n+11)u^4 +8(n+5)(n+9)(246+5n(n+14)u^2 
  \right. \nonumber \\
  &&
  \qquad \qquad \qquad 
  \left. 
    + 9 (n+6)^2 (n+8)^2 
  \right ) ,
  \nonumber \\
  &&
  \widetilde{Q}_5^{m=5}= 
  \frac{1}{4} \left(
    (n+3)(n+5)(n+12)(n+14)u^5 + 5(n+5)(n+12)(74+n(n+17))u^3 
  \right.  \nonumber \\
  &&
  \qquad \qquad \qquad 
  \left. 
    + 4(n+7)(n+8)(n+9)(n+10)u
  \right),
  \nonumber \\
  &&
  \widetilde{Q}_5^{m=6}= \frac{1}{64} \left( 
    64 (n+3) (n+5) (n+7) (n+13) (n+15) (n+17) u^6 
  \right. 
  \nonumber \\
  &&
  \qquad \qquad \qquad 
  +80(n+5)(n+7)(n+13)(n+15)(732+7n(n+20)) u^4 
  \nonumber \\
  &&
  \qquad \qquad \qquad 
  +4(n+7)(n+13)(2548800 +n(n+20)(51384+259 n (n+20)))u^2 
  \nonumber \\
  &&
  \qquad \qquad \qquad 
  \left. 
    +225 (n+8)^2(n+10)^2(n+12)^2   
  \right),
  \nonumber \\
  &&
  \widetilde{Q}_5^{m=7}= 
  \frac{1}{8} \left(
    (n+3) (n+5) (n+7) (n+16) (n+18) (n+20) u^7
  \right.
  \nonumber \\
  &&
  \qquad \qquad \qquad 
  +14 (n+5) (n+7) (n+16) (n+18) (n (n+23)+141) u^5
  \nonumber \\
  &&
  \qquad \qquad \qquad 
  +7 (n+7) (n+16) (n (n+23) (7 n (n+23)+1860)+123660) u^3
  \nonumber \\
  &&
  \qquad \qquad \qquad 
  \left.
    +36 (n+9) (n+10) (n+11) (n+12) (n+13) (n+14) u
  \right),
   \\
  &&
  \widetilde{Q}_5^{m=8}= 
  \frac{1}{256} \left(
    256 (n+3) (n+5) (n+7) (n+9) (n+17) (n+19) (n+21) (n+23) u^8
  \right.
  \nonumber \\
  &&
  \qquad \qquad \qquad 
  +1792 (n+5) (n+7) (n+9) (n+17) (n+19) (n+21) (3 n (n+26)+550) u^6
  \nonumber \\
  &&
  \qquad \qquad \qquad 
  +224(n+7) (n+9) (n+17) (n+19) (n (n+26) (141 n (n+26)+48544)
  \nonumber \\
  &&
  \qquad \qquad \qquad \qquad \qquad \qquad \qquad \qquad \qquad \qquad  
  +4185720) u^4
  \nonumber \\
  &&
  \qquad \qquad \qquad 
  +16 (n+9) (n+17) (n (n+26) (n (n+26) (3229 n(n+26)+1613162)
  \nonumber \\
  &&
  \qquad \qquad \qquad \qquad \qquad \qquad \qquad \qquad \qquad \qquad \qquad
  +268631440)+14910974400) u^2
  \nonumber \\
  &&
  \qquad \qquad \qquad 
  \left.
    +11025 (n+10)^2 (n+12)^2 (n+14)^2 (n+16)^2
  \right).\nonumber
\end{eqnarray}
For $\widetilde{Q}_7$ we have
\begin{eqnarray}
  \label{FirstQ7}
  &&
  \widetilde{Q}_7^{m=2}= 1,
  \nonumber \\
  &&
  \widetilde{Q}_7^{m=3}= u ,
  \nonumber \\
  &&
  \widetilde{Q}_7^{m=4}=
  n^2+4 (n+3) (n+11) u^2+14 n+60,
  \nonumber \\
  &&
  \widetilde{Q}_7^{m=5}= 
  (n+3) (n+14) u^3+(n (n+17)+90) u,
  \nonumber \\
  &&
  \widetilde{Q}_7^{m=6}=
  16 (n+3) (n+5) (n+15) (n+17) u^4
  \nonumber \\
  &&
  \qquad \qquad 
  +40 (n+5) (n+15) (n (n+20)+126) u^2
  \nonumber \\
  &&
  \qquad \qquad 
  +3 n (n+20) (3 n (n+20)+628)+100800,
  \nonumber \\
  &&
  \widetilde{Q}_7^{m=7}= 
  (n+3) (n+5) (n+18) (n+20) u^5
   \\
  &&
  \qquad \qquad 
  +5 (n+5) (n+18) (n (n+23)+168) u^3
  \nonumber \\
  &&
  \qquad \qquad 
  +4 (n(n+23) (n (n+23)+285)+20790) u,
  \nonumber \\
  &&
  \widetilde{Q}_7^{m=8}= 
  64 (n+3) (n+5) (n+7) (n+19) (n+21) (n+23) u^6
  \nonumber \\
  &&
  \qquad \qquad 
  +560 (n+5) (n+7) (n+19) (n+21) (n (n+26)+216) u^4
  \nonumber \\
  &&
  \qquad \qquad 
  +28 (n+7) (n+19) (n (n+26) (37 n(n+26)+13788)+1315440) u^2
  \nonumber \\
  &&
  \qquad \qquad 
  +45 (n (n+26) (n (n+26) (5 n (n+26)+2564)+439712)+25276160).
  \nonumber 
\end{eqnarray}

\section{Explicit formulae for the leading order wrapping correction}
\label{sec:Y-system}
 
The Y-system is a set of functional equations for the functions
$Y_{a,s}(u)$ defined on the fat-hook diagram associated with
$\mathfrak{psu}(2,2|4)$ which is a suitable $(a,s)$ grid described in
details in ~\cite{GKV,Gromov:2009tv}.
The anomalous dimension of a generic state is given by the TBA formula
\begin{equation}
  \label{eq:TBAenergy}
  E = \underbrace{\sum_{i} \epsilon_{1}(u_{4,i})}_{\rm asymptotic}+\underbrace{\sum_{a\ge 1}\int_{\mathbb R}\frac{du}{2\pi i}
    \frac{\partial \epsilon_{a}^{\star}}{\partial u}\,\log(1+Y^{\star}_{a,0}(u))}_{\rm wrapping\ W}.
\end{equation}
In this formula, the dispersion relation is
\begin{equation}
  \epsilon_{a}(u) = a+\frac{2\,i\,g}{x^{[a]}}-\frac{2\,i\,g}{x^{[-a]}},
\end{equation}
and the star means evaluation in the mirror kinematics~\footnote{ We
  recall that the physical and mirror branches of the Zhukowsky
  relation
  \begin{equation}
    x+\frac{1}{x} = \frac{u}{g},
  \end{equation}
  are
  \begin{equation}
    x_{\rm ph}(u) = \frac{1}{2}\left(\frac{u}{g}+\sqrt{\frac{u}{g}-2}\,\sqrt{\frac{u}{g}+2}\right), \qquad
    x_{\rm mir}(u) = \frac{1}{2}\left(\frac{u}{g}+i\,\sqrt{4-\frac{u^{2}}{g^{2}}}\right).
  \end{equation}
}.  The first term in $E$ is the sum of asymptotic one-magnon energies
and is the so-called asymptotic contribution to the anomalous
dimension. The second term is the wrapping correction.  The Bethe
roots $\{u_{4,i}\}$ are fixed by the exact Bethe equations (in
physical kinematics) $ Y_{1,0}(u_{4}) = -1 $.  Any solution of the
Y-system can be written in terms of a solution of the Hirota
integrable discrete equation.
For large $L$, (or small $g$) it can be shown that the Hirota equation
splits in two $\mathfrak{su}(2|2)_{\rm L, R}$ wings. One can have a
simultaneous finite large $L$ limit on both wings after a suitable
gauge transformation of the Hirota solution. Thus, we have
\begin{equation}
  \label{eq:Ysolution}
  Y_{a,0}(u) \simeq  \left(\frac{x^{[-a]}}{x^{[+a]}}\right)^{L}\,\frac{\Phi^{[-a]}}{\Phi^{[+a]}}\,
  T^{\rm L}_{a,1}\,T^{\rm R}_{a,1},
\end{equation}
where $\Phi$ is an arbitrary function and $T^{\rm L, R}_{a, 1}$ are
transfer matrices of the antisymmetric rectangular representations of
$\mathfrak{su}(2|2)_{\rm L, R}$. They are given explicitly by the
generating functional
\begin{eqnarray}
  \lefteqn{\sum_{a=0}^{\infty} (-1)^{a}\,T_{a,1}^{[1-a]}\,\ddb^{a} = \left(1-\frac{Q_{3}^{+}}{Q_{3}^{-}}\,\ddb\right)^{-1}\,
    \left(1-\frac{Q_{3}^{+}}{Q_{3}^{-}}\frac{Q_{2}^{--}}{Q_{2}}\frac{R^{(+)-}}{R^{(-)-}}\,\ddb\right)} && \\
  && 
  \qquad\times\left(1-\frac{Q_{2}^{++}}{Q_{2}}\frac{Q_{1}^{-}}{Q_{1}^{+}}\frac{R^{(+)-}}{R^{(-)-}}
    \,\ddb\right)\,
  \left(1-\frac{Q^{-}_{1}}{Q_{1}^{+}}\frac{B^{(+)+}}{B^{(-)+}}\frac{R^{(+)-}}{R^{(-)-}}\ddb\right)^{-1},
  \nonumber
\end{eqnarray}
where $\ddb = e^{-i\partial_{u}}$ and 
\begin{equation} 
  R_{}^{(\pm)} =
  \prod_{i=1}^{K_{4}} \frac{x(u)-x_{4,i}^{\mp}}{(x^{\mp}_{4,i})^{1/2}},
  \qquad B_{}^{(\pm)} = \prod_{i=1}^{K_{4}}
  \frac{\frac{1}{x(u)}-x_{4,i}^{\mp}}{(x^{\mp}_{4,i})^{1/2}}.
\end{equation}
The function $\Phi$ has been determined  in \cite{GKV} and reads
\begin{equation}
  \frac{\Phi^{-}}{\Phi^{+}} = \sigma^{2}\,\frac{B^{(+)+}\,R^{(-)-}}{B^{(-)-}\,R^{(+)+}}\,
  \frac{B^{+}_{1, \rm L}\,B^{-}_{3, \rm L}}{B^{-}_{1, \rm L}\,B^{+}_{3, \rm L}}\,
  \frac{B^{+}_{1, \rm R}\,B^{-}_{3, \rm R}}{B^{-}_{1, \rm R}\,B^{+}_{3, \rm R}},
\end{equation}
where $\sigma$ is the dressing phase
. At weak coupling, evaluating the various terms at leading order in
the mirror dynamics, the wrapping correction (second term in the
r.h.s. of (\ref{eq:TBAenergy})) is simply given by the expression
\begin{equation}
  \label{wrapping}
  W = -\frac{1}{\pi}\sum_{a=1}^{\infty}\int_{\mathbb R}du\,Y_{a,0}^{\star}.
\end{equation}

\subsection{Explicit formulae for the computation of $Y^{\star}_{a, 0}$}

In the following, we shall need a compact efficient formula for the
evaluation of $Y^{\star}_{a, 0}$.  According to (\ref{eq:Ysolution}),
we need the contribution from the dispersion (ratio of $x^{\pm}$), the
fusion of scalar factors ($\Phi$ terms), and the $\mathfrak{su}(2|2)$
transfer matrices.  After a straightforward computation we obtain:

\subsection*{Transfer matrices}
Using the relations, valid at leading order in the coupling constant
\begin{eqnarray}
  \frac{R^{(+)}}{R^{(-)}}&=&\frac{Q_4^{[+]}}{Q_4^{[-]}} \left(1+g^2\frac{i c}{u} + \mathcal{O}(g^4)\right),
  \nonumber \\
  \frac{B^{(+)}}{B^{(-)}}&=&\left(1-g^2\frac{i c}{u} + \mathcal{O}(g^4)\right),
  \nonumber \\
  c & = & \sum_j \frac{1}{u_{4,j}^+ u_{4,j}^-} = i \left(\log\left(Q_4\right)\right)'\bigg|_{u=-\frac{i}{2}}^{u=+\frac{i}{2}}.
\end{eqnarray}
we get the following expression for the transfer matrices $T^*_{a,1}$
in mirror dynamics:
\begin{eqnarray}
  \label{QuickTRight}
  T^*_{a,1} &=&
  (-1)^{a+1} \frac{Q_5^{[a]} Q_7^{[-a]}}{Q_4^{[1-a]}}  
  \mathop{\sum^{a-1}_{k=1-a}}_{\Delta k=2}
   \frac{Q_4^{[k]}}{Q_6^{[k]}} \left(\frac{Q_6^{[k+2]}-Q_6^{[k]}}{Q_5^{[k+1]} Q_7^{[k+1]}}
    + \frac{Q_6^{[k-2]}-Q_6^{[k]}} {Q_5^{[k-1]} Q_7^{[k-1]}}\right) +\mathcal{O}(g^2).
  \nonumber \\
\end{eqnarray}
We remark that this expression is valid for {\bf any distributions of
  roots} on Dynkin diagrams like the one of the right wing of picture
(\ref{eq:dynkin3}). So, for example, taking the expression for
$\widetilde{Q}_{5,m=2}$, $Q_{6,m=2}$, $\widetilde{Q}_{7,m=2}$ from
equations (\ref{FirstQ5}), (\ref{FirstQ6}) and (\ref{FirstQ7}), we get
back to eq. (4.22) of \cite{BMR11}. If $Q_6$ is trivial, i.e. $Q_6=1$,
formula (\ref{QuickTRight}) shows that the transfer matrix is
$\mathcal{O}(g^2)$. 

To compute the wrapping corrections we can use formula
(\ref{QuickTRight}) for the transfer matrices of the right wing of the
diagram (\ref{eq:dynkin3}), while for the $\mathcal{O}(g^2)$ left wing
we use
\begin{eqnarray}
  \label{QuickTLeft}
  T^{*,L}_{a,1} &=&
  i c g^2 \frac{(-1)^{a+1}}{Q_4^{[1-a]}} 
  \mathop{\sum^{a}_{k=-a}}_{\Delta k =2}
\frac{Q_4^{[-1-k]} - Q_4^{[1-k]}}{u-i\frac{k}{2}}\bigg|_{Q_4^{[-1-a]},Q_4^{[-1-a]}\rightarrow 0}  
  +\mathcal{O}(g^4).
\end{eqnarray}

\subsection*{Dispersion relation}
This is the universal factor
\begin{equation}
  \label{dispersion}
  \left(\frac{4 g^2}{ a^2 + 4 u^2}\right)^3.
\end{equation}

\subsection*{Fusion scalar factor}
From the relation
\begin{equation}
  \label{FusionGen}
  \frac{\Phi^-}{\Phi^+} = 
  \sigma^2 ~
  \frac{B^{(+)+} R^{(-)-}}{B^{(-)-} R^{(+)+}} ~
  \frac{B_1^{+} B_3^{-}}{B_1^{-} B_3^{+}} ~
  \frac{B_7^{+} B_5^{-}}{B_7^{-} B_5^{+}} ,
\end{equation}
the following formula follows
\begin{eqnarray}
  \label{Fusion}
  \Phi^*_a = \left[Q_4^+(0)\right]^2 \frac{Q_4^{[1-a]}}{Q_4^{[-1-a]} Q_4^{[a-1]} Q_4^{[a+1]}} ~ 
  \frac{Q_5^{[-a]}}{Q_5(0)} ~
  \frac{Q_7(0)}{Q_7^{[-a]}}.
\end{eqnarray}
This formula is valid for even $Q_4$, $Q_5$ and $Q_7$, i.e. for even
values of $n$ and $m$. For $m$ odd, the ratio $Q_7/Q_{5}$ is
indeterminate at $u=0$, but has a smooth limit for $u\to 0$.
  
\section{Large $n$ expansion: The algorithm}
\label{sec:algorithm}

The wrapping correction can be computed by summing the residues of the
$Y_a$--functions at $u=\frac{i a}{2}$. The precise relation is
\begin{equation} 
  \label{wresidues} 
  W =   -\frac{1}{\pi}\sum_{a=1}^{\infty}\int_{\mathbb R}du\,Y_{a,0}^{\star}
  = - 2 i \sum_{a=1}^{\infty} \mathop{\textrm{Res}}_{u=i\frac{a}{2}}
  Y_{a,0}^{\star}.  
\end{equation} 
The physical reason of this property, that we explicitly checked for
all the cases we are interested in, is presumably the same as in the
Konishi case discussed in \cite{Bajnok:2008qj}. The pole at
$u=\frac{ia}{2}$ is of kinematical origin and does not depend on the
scattering matrix. Instead, other poles are determined by the dynamics
and correspond to $\mu$ terms in the L\"uscher approach to wrapping
corrections. It is expected that such terms are absent in the weakly
coupled limit \cite{Bajnok:2008bm}.

\medskip Since we are interested in the large spin limit of $W$, we
can attempt to exchange this limit with the sum over the intermediate
virtual states in the r.h.s of Eq.~(\ref{wresidues}). This possibility
is supported by the fact that the large $n$ structure perfectly
matches the exact result in all known cases in the $\mathfrak{sl}(2)$
sector of $\mathcal{N}=4$ SYM theory, in its $\beta$-deformed version
and in ABJM theory as shown in \cite{BLM}.

\medskip In practice, one evaluates the above residue at fixed $a=1,
2, \dots$ without assigning $n$ and then taking the limit over it in
two steps: The dependence on $n$, in fact, comes from the polynomials
$Q_{4}$, its derivatives (which are written in terms of the basic
hypergeometric function $F_{n,m}$ defined in (\ref{hypergeo})) and
from the explicit $n$-dependent coefficients of the other Baxter
polynomials.  At this point one can use the Baxter equation to shift
the argument of $F_{n,m}$ to some minimal value and take the large $n$
limit on the coefficients. This gives a first expansion containing
various derivatives of the logarithm of $F_{n,m}$ which in turn can be
systematically computed as explained in Appendix~\ref{app:exp} or by
means of the method explained in \cite{Beccaria:2007uj}. The outcome
of this procedure are sequences of rational numbers being the
$a$-dependent coefficients of the large $n$ expansion of
$\mathop{\textrm{Res}}_{u=i\frac{a}{2}} Y_{a,0}^{\star}$. These
sequences turn out to be rather simple rational functions which are
easily identified and summed over $a$.

\medskip In the following, we first apply this strategy to the case
$m=2$ reproducing the known results for 3-gluon operators.  Then, we
 move to unexplored cases $m>2$ for which we  provide new
asymptotic expansions for the wrapping correction.

\subsection{$m=2$, checking 3-gluon operators}

Let $L(u)$ be the logarithm of the basic hypergeometric function
\begin{equation}
  L(u) = \log F_{n,m=2}(u),
\end{equation}
and $R_{a}$ be the residue
\begin{equation}
  g^{8}\,R_{a}(n) = \mathop{\mbox{Res}}_{u=\frac{ia}{2}} Y_{a,0}^{\star}(u).
\end{equation}
We find the following explicit results for the first residues expanded
at first order for large $n \equiv 1/\epsilon$
\begin{eqnarray}
  R_{1} &=& 
  \frac{160}{81} \epsilon ^2 \left(L'\left(\frac{i}{2}\right)-5 i\right) 
  \left(76 + 9 L''\left(\frac{i}{2}\right)\right) +O\left(\epsilon^3\right),
  \nonumber \\
  R_{a\ge 2} 
  &=& 
  32 \epsilon ^2 \left(L'\left(\frac{i}{2}\right)-5 i\right) 
  \left(f_1\left(a\right)+f_2\left(a\right) L''\left(\frac{i}{2}\right)\right)+O\left(\epsilon^3\right).
\end{eqnarray}
Notice that the whole dependence on the spin $n$ is inside the
derivatives of $L(u)$ evaluated at special points.  Instead, the
dependence on the label of intermediate virtual states $a$ is in the
coefficient functions $f_{1,2}(a)$.  In principle these functions
could be very non trivial. In our case, we find that they are rather
simple rational functions precisely as in other cases analyzed in
\cite{BLM}. In particular, we find
\begin{eqnarray}
  f_{1}(a) &=&  \frac{\left(192 a^{10}-960 a^9+2640 a^8-4800 a^7+4916 a^6-1980 a^5-405 a^4+430 a^3+12
      a^2-45 a+9\right)}{(a-1)^3 \,a^3 \,(2 a-3)^3 (2 a-1)^3 (2 a+1)^3},  
  \nonumber \\
  f_{2}(a) &=&
  \frac{1}{(a-1)\, a\, (2 a-3) (2 a-1) (2 a+1)}.
\end{eqnarray}
The derivatives $L^{(n)}\left(\frac{i}{2}\right)$ are computed in
Appendix A.  Summing over $a$ and in terms of $\overline n\equiv
\frac{1}{2}\,e^{\gamma_{E}}\,n$ ($\gamma_{E}$ is the Euler number), we
find
\begin{eqnarray}
  && \sum_{a=1}^{\infty}R_{a}= -\frac{256\,i}{3} (3\zeta_{3}-1)\, \frac{\log\overline n+1}{n^2}+
  \mathcal O\left(\frac{\log \bar n}{n^{3}}\right).
  \nonumber 
\end{eqnarray}
This result is in perfect agreement with the large $n$ expansion of
the results of \cite{BMR11}. This provides the validity of the
computational method we are using. We now apply this same method to
cases with $m>2$.

\subsection{$m=3$}

As in the previous case, let $L(u)$ be the logarithm of the basic hypergeometric
function 
\begin{equation}
  L(u) = \log F_{n,m=3}(u),
\end{equation}
and $R_{a}$ be the residue
\begin{equation}
  g^{8}\,R_{a}(n) =  \mathop{\mbox{Res}}_{u=\frac{ia}{2}} Y_{a,0}^{\star}(u).
\end{equation}
We find the following explicit results for the first three residues
expanded at large $n \equiv 1/\epsilon$
\begin{eqnarray}
  R_{1} &=& \frac{9}{2} \epsilon ^2 \left(2 L'(i)-11 i\right) \left(6+L''(i)\right)-\frac{99}{2} \epsilon ^3 \left(\left(2 L'(i)-11 i\right) \left(6+L''(i)\right)\right)+\nonumber \\
  &&+\frac{3}{8} \epsilon ^4 \left(-11624 i L''(i)+12590
    L'(i)+2096 L'(i) L''(i)-69821 i\right)+O\left(\epsilon ^5\right), \\
  R_{2} &=& \frac{1}{36} \epsilon ^2 \left(2 L'(i)-11 i\right) \left(191+18 L''(i)\right)-\frac{11}{36} \epsilon ^3 \left(\left(2 L'(i)-11 i\right) \left(191+18 L''(i)\right)\right)+\nonumber \\
  && +\frac{1}{216} \epsilon ^4 \left(-84816 i
    L''(i)+177886 L'(i)+15264 L'(i) L''(i)-987541 i\right)+O\left(\epsilon ^5\right), \nonumber\\
  R_{3} &=& -\frac{35}{288} \epsilon ^2 \left(2 L'(i)-11 i\right)+\frac{385}{288} \epsilon ^3 \left(2 L'(i)-11 i\right)+\nonumber \\
  &&+\frac{1}{576} \epsilon ^4 \left(-8448 i L''(i)-1758 L'(i)+1536 L'(i) L''(i)+10229 i\right)+O\left(\epsilon
    ^5\right).\nonumber
\end{eqnarray}
For $a\ge 4$, we find instead
\begin{eqnarray}
  R_{a\ge 4} &=& (\epsilon^{2}-11\,\epsilon^{3})\,\left[L'(i)-\frac{11i}{2}\right]\,f_{1}(a) + \\
  &&  +\epsilon^{4}[L'(i)\,f_{2}(a)+f_{3}(a)]+\mathcal O(\epsilon^{5}),
\end{eqnarray}
where again we can match the functions $f$'s with rational functions
\begin{eqnarray}
  f_{1}(a) &=&  -\frac{12 (2 a-1) \left(3 a^2-3 a-4\right)}{(a-2)^2 (a-1)^3 a^3 (a+1)^2},\\
  f_{2}(a) &=& -\frac{12 \left(20 a^8+426 a^7-4399 a^6+12288 a^5-6156 a^4-19036 a^3+17131 a^2+8538 a-6108\right)}
  {(a-3)^2 (a-2)^3 (a-1)^3 a^3 (a+1)^3},\nonumber\\
  f_{3}(a) &=& \frac{6 i \left(220 a^8+4734 a^7-48797 a^6+136256 a^5-68196 a^4-211220 a^3+190033 a^2+94734 a-67764\right)}
  {(a-3)^2 (a-2)^3 (a-1)^3 a^3 (a+1)^3}.\nonumber 
\end{eqnarray}
The derivatives $L^{(n)}(i)$ are computed in Appendix A. Summing over
$a$, we find the final result (setting now $\overline n\equiv e^{\gamma_{E}}\,n$)
\begin{eqnarray}
  && \sum_{a=1}^{\infty}R_{a}= -\frac{2\,i}{3}(2\log\overline n+1)(-3+5\pi^{2}+36\zeta_{3})\,\frac{1}{n^{2}}+ \\
  && \frac{44\,i}{3}\,\log\overline n\,(-3+5\pi^{2}+36\zeta_{3})\,\frac{1}{n^{3}}+\nonumber \\
  &&\left[
    -\frac{4\,i}{9}(-615+1289\pi^{2}+9108\zeta_{3})\,\log\overline n+\frac{i}{9}(-1527
    +2017\pi^{2}+14868\zeta_{3})
  \right]\,\frac{1}{n^{4}}+\dots\nonumber
\end{eqnarray}
A comparison with a numerical estimate of the (imaginary part of the)
sum of the residues is shown in the following table 
\begin{equation}
  \begin{array}{ccccccc}
    n & \mbox{estimate} & ({\rm LO} & {\rm NLO} & {\rm NNLO}) & \mbox{full expansion} & \mbox{diff \%}\\
    \hline
    10 & -1.857411 & (-4.038730 & 3.785374 & - 2.548066) & -2.801422  & 51 \%\\
    30 & -0.438781 & (-0.594614 & 0.193683 & -0.045355) &  -0.446286  & 1.7 \%\\
    50 & -0.197270 & (-0.238478 & 0.047207 & - 0.006716)  & -0.197986 & 0.36 \%
  \end{array}
\end{equation}

\subsection{$m=4$}

Again, let $L(u)$ be the logarithm of the basic hypergeometric function
\begin{equation}
  L(u) = \log F_{n,m=4}(u),
\end{equation}
and  $R_{a}$ be the residue
\begin{equation}
  g^{8}\,R_{a}(n) =  \mathop{\mbox{Res}}_{u=\frac{ia}{2}} Y_{a,0}^{\star}(u).
\end{equation}
We find the following explicit results for the first two residues
expanded at large $n \equiv 1/\epsilon$
\begin{eqnarray}
  R_{1} &=&\epsilon ^2 \left(-\frac{9632}{225} i L''\left(\frac{i}{2}\right)+\frac{942928 L'\left(\frac{i}{2}\right)}{16875}+\frac{448}{75} L'\left(\frac{i}{2}\right) L''\left(\frac{i}{2}\right)-\frac{20272952
      i}{50625}\right)+\\
  && +\epsilon ^3 \left(\frac{134848}{225} i L''\left(\frac{i}{2}\right)-\frac{13200992 L'\left(\frac{i}{2}\right)}{16875}-\frac{6272}{75} L'\left(\frac{i}{2}\right)
    L''\left(\frac{i}{2}\right)+\frac{283821328 i}{50625}\right)+\nonumber\\
  &&+\epsilon ^4 \left(-\frac{7420192 i L''\left(\frac{i}{2}\right)}{1125}+\frac{28672}{75} L'\left(\frac{i}{2}\right)^3-\frac{917504}{225} i
    L'\left(\frac{i}{2}\right)^2-\frac{328511984 L'\left(\frac{i}{2}\right)}{84375}+\right. \nonumber \\
  && \left. +\frac{288448}{375} L'\left(\frac{i}{2}\right) L''\left(\frac{i}{2}\right)-\frac{12897148504 i}{253125}\right)+\mathcal O\left(\epsilon
    ^5\right),\nonumber
\end{eqnarray}
\begin{eqnarray}
  R_{2} &=&\epsilon ^2 \left(-\frac{688}{105} i L''\left(\frac{i}{2}\right)+\frac{4510928 L'\left(\frac{i}{2}\right)}{385875}+\frac{32}{35} L'\left(\frac{i}{2}\right) L''\left(\frac{i}{2}\right)-\frac{96984952
      i}{1157625}\right)+\\
  && +\epsilon ^3 \left(\frac{1376}{15} i L''\left(\frac{i}{2}\right)-\frac{9021856 L'\left(\frac{i}{2}\right)}{55125}-\frac{64}{5} L'\left(\frac{i}{2}\right)
    L''\left(\frac{i}{2}\right)+\frac{193969904 i}{165375}\right)+\nonumber \\
  &&+\epsilon ^4 \left(-\frac{48512752 i L''\left(\frac{i}{2}\right)}{55125}+\frac{2048}{35} L'\left(\frac{i}{2}\right)^3-\frac{65536}{105} i
    L'\left(\frac{i}{2}\right)^2-\frac{10476726928 L'\left(\frac{i}{2}\right)}{28940625}+\right. \nonumber\\
  && \left. +\frac{1831328 L'\left(\frac{i}{2}\right) L''\left(\frac{i}{2}\right)}{18375}-\frac{872054313448
      i}{86821875}\right)+\mathcal O\left(\epsilon ^5\right).\nonumber
\end{eqnarray}
For $a\ge 3$, we find the same general structure
\begin{eqnarray}
  R_{a\ge 3} &=& 
  \epsilon^{2}\,\left(
    f_{2,0}(a)+f_{2,1}(a)\, L'\left(\frac{i}{2}\right)+f_{2,2}(a)\, L''\left(\frac{i}{2}\right)+
    f_{2,12}(a)\, L'\left(\frac{i}{2}\right)L''\left(\frac{i}{2}\right)
  \right)+\nonumber\\
  &&+\epsilon^{3}\,\left(
    f_{3,0}(a)+f_{3,1}(a)\, L'\left(\frac{i}{2}\right)+f_{3,2}(a)\, L''\left(\frac{i}{2}\right)+
    f_{3,12}(a)\, L'\left(\frac{i}{2}\right)L''\left(\frac{i}{2}\right)
  \right)+\nonumber \\
  &&+\epsilon^{4}\,\left(
    f_{4,0}(a)+f_{4,1}(a)\, L'\left(\frac{i}{2}\right)+f_{4,11}(a)\, L'\left(\frac{i}{2}\right)^{2}+f_{4,111}(a)\, L'\left(\frac{i}{2}\right)^{3}+\right. \nonumber \\
  &&\left. +f_{4,2}(a)\, L''\left(\frac{i}{2}\right)+
    f_{4,12}(a)\, L'\left(\frac{i}{2}\right)L''\left(\frac{i}{2}\right)
  \right)+\dots,\nonumber
\end{eqnarray}
where
\begin{eqnarray}
  f_{2,0}(a) &=&  \frac{344\, i}{(a-1)^3 a^3 (2 a-5)^3 (2 a-3)^3 (2 a-1)^3 (2 a+1)^3 (2
    a+3)^3} \times \nonumber \\ 
  && \quad \left(15360\, a^{14}-107520\, a^{13}+302848\,
    a^{12}-419328\,
    a^{11}-61824\, a^{10}+ \right.  \nonumber \\
  &&  \quad   \quad   \left. 1590400\, a^9-2375328\, a^8-260736\, a^7+3289132\,
    a^6-2275140\, a^5- \right.   \nonumber \\ 
  &&  \quad   \quad \quad \left. 26229\,  a^4+382662\, a^3-20277\, a^2-34020\,
    a+8100\right),\\{}\nonumber\\
  f_{2,1}(a) &=&\frac{6\, i}{43}\,\,  f_{2,0}(a), \\{}\nonumber\\
  f_{2,2}(a) &=&\frac{1376\, i}{(a-1) a (2 a-5) (2 a-3) (2 a-1) (2 a+1) (2 a+3)},\\{}\nonumber\\
  f_{2,12}(a) &=& \frac{6\, i}{43}\,\,  f_{2,2}(a),\\ {}\nonumber\\
  f_{3,0}(a) &=& - 14\,\, f_{2,0}(a),\qquad f_{3,1}(a) = \frac{6\, i}{43}\,\, f_{3,0}(a), \\{}\nonumber\\
  f_{3,2}(a) &=& - 14\,\, f_{2,2}(a), \qquad f_{3,1,2}(a) =  \frac{6\,
    i}{43}\,\, f_{3,2}(a), \\{}\nonumber\\
  f_{4,0}(a) &=& \frac{8 i}{(a-2)^3 (a-1)^3 a^3 (2 a-7)^3 (2 a-5)^3 (2 a-3)^3 (2
    a-1)^3 (2 a+1)^4 (2 a+3)^4}\times \nonumber \\ 
  &&\quad \left(105676800\, a^{23}+193789952\, a^{22}-30633238528\,
    a^{21}+374751035392\, a^{20} - \right. \nonumber \\  
  &&\quad \left.  2338797209600\, a^{19}+9042207928320\,
      a^{18}-21933773253120\, a^{17}+  \right. \nonumber \\
  &&\quad \left. 25092446111744\, a^{16}+  32261701882496\, a^{15}-181183264537856\,
    a^{14}+ \right. \nonumber \\
  &&\quad \left. 259510085417632\,  a^{13}+62844522906624\, a^{12}-666497213030424\,
    a^{11}+ \right. \nonumber \\
  &&\quad \left. 681006561227024\, a^{10}+220531002799826\, a^9-879621870442904\,
    a^8+  \right. \nonumber \\
  &&\quad \left. 464871876325207\, a^7+160461344177928\, a^6-197687865980727\, a^5
  \right. \nonumber \\
  &&\quad \left.  + 10634069596050\, a^4+27492240914352\, a^3-3490888295808\, a^2 - \right. \nonumber \\
  &&\quad \left. 1670217479280\, a+429309266400\right),\\{}\nonumber\\
  f_{4,1}(a) &=& -\frac{16}{(a-2)^3 (a-1)^3 a^3
    (2 a-7)^3 (2 a-5)^3 (2 a-3)^3 (2 a-1)^3 (2 a+1)^4 (2 a+3)^4}\times \nonumber \\ 
  &&\quad \left(7372800\, a^{23}-655032320\, a^{22}+12236677120\,
    a^{21}-  \right. \nonumber \\  
  &&\quad \left.
    103330938880\, a^{20}+446158370816\, a^{19}-770280591360\,
    a^{18}-   \right. \nonumber \\  
  &&\quad \left. 
    1416142109184\, a^{17}+9605107320832\, a^{16}-15447524759936\,
    a^{15}-  \right. \nonumber \\  
  &&\quad \left. 
    8228184007936\, a^{14}+60190240489376\, a^{13}-59275890588672\,
    a^{12}-  \right. \nonumber \\  
  &&\quad \left. 
    47305035162840\, a^{11}+130190626486672\, a^{10}-49532653784990\,
    a^9- \right. \nonumber \\  
  &&\quad \left. 
    75968166506584\, a^8+70367565767567\, a^7+3329446699848\,
    a^6- \right. \nonumber \\  
  &&\quad \left. 
    21692142654831\,  a^5+3791318445090\, a^4+2416824867888\, a^3-  \right. \nonumber \\  
  &&\quad \left. 
    531904185216\, a^2-96835480560\, a+24890392800\right),\\{}\nonumber\\
  f_{4,11}(a) &=& \frac{131072\, i}{(a-1) a (2 a-5) (2 a-3) (2 a-1) (2 a+1) (2 a+3)},\\{}\nonumber\\
  f_{4,111}(a) &=&-\frac{12288}{(a-1) a (2 a-5) (2 a-3) (2 a-1) (2 a+1) (2
    a+3)},\\{}\nonumber\\
  f_{4,2}(a) &=&
  \frac{32 i \left(1720\, a^5+56864\, a^4-203410\, a^3-163652\, a^2+521499\,
    a+270414\right)}{(a-2) (a-1) a (2 a-7) (2 a-5) (2 a-3) (2 a-1) (2
   a+1)^2 (2 a+3)^2},\\{}\nonumber\\
  f_{4,12}(a) &=&
-\frac{192 \left(40\, a^5+1120\, a^4-4022\, a^3-3148\, a^2+10129\, a+5226\right)}{(a-2)
  (a-1) a (2 a-7) (2 a-5) (2 a-3) (2 a-1) (2 a+1)^2 (2
   a+3)^2}.
\end{eqnarray}
The derivatives $L^{(n)}\left(\frac{i}{2}\right)$ are computed in
Appendix A.  Summing over $a$, we find the final result ($\overline
n\equiv \frac{1}{2}\,e^{\gamma_{E}}\,n$)
\begin{eqnarray}
  && \sum_{a=1}^{\infty}R_{a}= -\frac{512\, i}{1215} \left(3 \log\overline n + 4\right)\left(-32+81\zeta_{3}\right)\,\frac{1}{n^{2}}+
  \nonumber\\
  && \frac{3584\, i}{1215} \left(6 \log\overline n + 5\right)\left(-32+81\zeta_{3}\right)\,\frac{1}{n^{3}}+ \\
  && \frac{512\, i}{8505} \left(135 (1971 \zeta_3-760) \log\overline n+143289 \zeta_3-53248\right) \,\frac{1}{n^{4}}
  \nonumber
\end{eqnarray}
Again, we can present a numerical table showing the accuracy of the computed asymptotic expansion
\begin{equation}
  \begin{array}{ccccccc}
    n & \mbox{estimate} & ({\rm LO} & {\rm NLO} & {\rm NNLO}) & \mbox{full expansion} & \mbox{diff \%}\\
    \hline
    10 & -1.201460 & ( - 2.908787 & 3.493848 &  - 3.576129 ) &  -2.991068 & 149 \% \\
    30 & -0.293765 & (- 0.424071 & 0.176476 & - 0.061888) &  -0.309484 & 5.4\% \\
    50 & -0.134246 & (- 0.169551 & 0.042847 & - 0.009090 ) & - 0.135794 & 1.2 \%
  \end{array}
\end{equation}

\section{Summary and a reciprocity conjecture}
\label{sec:summary}

In summary, our results for the large spin expansion of the leading
order wrapping correction at $m=2,3,4$ are (we set here $\bar n =
e^{\gamma_{\rm E}}\,n$ for all $m$)
\begin{eqnarray}
  g^{-8}\,W_{n,m=2} &=& -\frac{512}{3}\, (3\,\zeta_{3}-1)\,\frac{3\,\log\frac{\bar n}{2}+1}{n^{2}}
  +\frac{2048}{3}\, (3\,\zeta_{3}-1)\,\frac{2\,\log\frac{\bar n}{2}+1}{n^{3}}+\\
  && -\frac{1536}{5}\,(77\,\zeta_{3}-24)\,
  \frac{\log{\frac{\bar n}{2}}}{n^{4}}+\dots, \nonumber \\
  g^{-8}\,W_{n,m=3} &=& -\frac{4}{3}\, (36\,\zeta_{3}+5\pi^{2}-3)\,\frac{2\,\log{\bar n}+1}{n^{2}}
  +\frac{88}{3}\, (36\,\zeta_{3}+5\pi^{2}-3)\,\frac{\log\bar n}{n^{3}}+\\
  && -\frac{2}{9\,n^{4}}\left[
    4\,(9108\,\zeta_{3}+1289\pi^{2}-615)\,\log\bar n-14868\,\zeta_{3}-2017\pi^{2}+1527
  \right]+\dots, \nonumber\\
  g^{-8}\,W_{n,m=4} &=& -\frac{1024}{1215}\, (81\,\zeta_{3}-32)\,\frac{3\,\log{\frac{\bar n}{2}}+4}{n^{2}}
  +\frac{7168}{1215}\, (81\,\zeta_{3}-32)\,\frac{6\,\log\frac{\bar n}{2}+5}{n^{3}}+\\
  && -\frac{1024}{8505\,n^{4}}\left[
    138240\,(1971\,\zeta_{3}-760)\,\log\frac{\bar n}{2}+143289\,\zeta_{3}-53248\right]+\dots. \nonumber
\end{eqnarray}
Following the general idea of \cite{reciprocity} (see for instance the
review \cite{Beccaria:2010tb} for its many tests in AdS/CFT), we are
led to rewrite the above large $n$ expansions in terms of the quantity
\begin{equation}
  \mathcal J^{2}_{m} = n\,(n+a_{m}).
\end{equation}
The possible vanishing of odd terms $1/\mathcal J^{2k+1}$ is linked to
the Gribov-Lipatov reciprocity and allows to interpret $\mathcal J$ as
the Casimir of a suitable additional symmetry of anomalous dimensions.
From previous experience, it can be expected such reciprocity
relations to hold not only for the full anomalous dimension, but also
separately for the leading order wrapping correction. It turns out
that the coefficients of the two odd terms $1/\mathcal J^{3}$ and
$\log\mathcal J/\mathcal J^{3}$ indeed vanish for the choice \be a_{2,
  3, 4} = 8, \ 11, \ 14.  \ee This is not completely trivial since we
have one parameters and two structures.  It is tempting to conjecture
the simple relation $a_{m} = 3\,m+2$ and to claim that reciprocity in
the above sense holds for the full anomalous dimension as well. This
remark could help in the task of finding a closed expression for the
asymptotic anomalous dimensions which is currently unavailable beyond
one loop. For completeness, we report the expansion of wrapping in
terms of $\mathcal J_{m} = n\,(n+3\,m+2)$
\begin{eqnarray} 
  g^{-8}\,W_{n,m=2} &=&
  -\frac{256}{3\mathcal J^{2}_{2}}\,(3\,\zeta_{3}-1)\,
  \left(\log\frac{\bar{\mathcal J_{2}}^{2}}{4}+2\right)+\\
  &&+\frac{256}{15\mathcal J^{4}_{2}}\,\left[
    (267\,\zeta_{3}-104)\,\log\frac{\bar{\mathcal J_{2}}^{2}}{4}+480\,\zeta_{3}-160\right],\nonumber \\
  g^{-8}\,W_{n,m=3} &=& -\frac{4}{3\mathcal
    J^{2}_{3}}\,(36\,\zeta_{3}+5\pi^{2}-3)\,
  \left(\log{\bar{\mathcal J_{3}}^{2}}+1\right)+\\
  &&+\frac{4}{9\mathcal J^{4}_{3}}\,\left[
    2\,(1980\,\zeta_{3}+263\pi^{2}-237)\,\log{\bar{\mathcal J_{3}}^{2}}+900\,\zeta_{3}+101\pi^{2}-219\right],\nonumber \\
  g^{-8}\,W_{n,m=4} &=& -\frac{512}{1215\mathcal
    J^{2}_{4}}\,(81\,\zeta_{3}-32)\,
  \left(3\,\log\frac{\bar{\mathcal J_{4}}^{2}}{4}+8\right)+\\
  &&+\frac{512}{8505\mathcal J^{4}_{4}}\,\left[
    (67311\,\zeta_{3}-29112)\,\log\frac{\bar{\mathcal
        J}^{2}_{4}}{4}+102384\,\zeta_{3}-47168\right].\nonumber 
\end{eqnarray}

\section{Conclusions}

In this paper we have applied a simple algorithm to derive the large
spin expansion of the leading order wrapping correction to a class of
twist operators introduced by S. Zieme, A. Rej and L. Freyhult in
\cite{Freyhult:2009fc}.  Our analysis extends previous work on simple
$\mathfrak{sl}(2)$-like rank one classes of states in $\beta$-deformed
or ABJM theories. We could easily obtain accurate asymptotic
expansions for various special cases.  This analytic results can be
used to claim the correct scaling behaviour of the wrapping
correction, but also to explore other interesting properties like
reciprocity constraint. In principle, our analysis could be helpful in
a possible attempt to derive the currently unavailable explicit
expression of the asymptotic anomalous dimension beyond one-loop.

\section*{Acknowledgments}

We thank M. Staudacher, Stefan Zieme  and Nikolay Gromov for helpful  discussions.

\appendix
\section{One-loop explicit Baxter polynomials}
\label{app:oneloop-solution}

The general solutions to the 1-loop Bethe equations (\ref{Bethe-eq})
are given by the following Baxter polynomials \cite{Freyhult:2009fc}
\begin{eqnarray}
  \label{Q4}
  Q_4(u) & = & \sum_{k=0}^{m} \left(-1\right)^k \binom{m}{k} 
  \prod_{j=1}^{k}   \left(u -i \frac{2j-1}{2}\right)^3
  \prod_{j=1}^{m-k} \left( u +i \frac{2j-1}{2}\right)^3
  F_{n,m}\left(u+i\frac{m-2k}{2}\right),\nonumber
  \\
  \label{Q5}
  Q_5(u) & = & \sum_{k=0}^{m-1} \left(-1\right)^k \binom{m-1}{k} 
  \prod_{j=1}^{k}   \left(u -i j\right)^3
  \prod_{j=1}^{m-1-k} \left( u +i j\right)^3
  F_{n,m}\left(u+i\frac{m-1-2k}{2}\right),
  \nonumber \\
  \label{Q6}
  Q_6(u) & = &  \prod_{k=0}^{m-2} f_k\left( u +i \frac{k}{2}\right)
   \\
  &&
  +\sum_{r=1}^{m-1} (-1)^r \sum_{j_1=0}^{m-2} \sum_{j_2=0}^{j_1-1} \cdots \sum_{j_r=0}^{j_{r-1}-1} 
  \prod_{s=1}^{r}\tilde{f}_{j_s}\left(u+ i \frac{j_s-2(r-s)}{2} \right)
  \prod_{k=0}^{j_r -1} f_{k}\left(u+ i \frac{k}{2} \right)
  \nonumber \\
  &&
  \times \prod_{s=2}^{r} \prod_{k=j_s +1}^{j_{s-1} -1} f_{k}\left(u+ i \frac{k-2(r-s+1)}{2} \right)
  \prod_{k=j_1 +1}^{m-2} f_{k}\left(u+ i \frac{k-2r}{2} \right),\nonumber
\end{eqnarray}
where the hypergeometric function
\begin{equation}
  \label{hypergeo}
  F_{n,m}(u)={}_{4}F_{3}\left(
    \left.
      \begin{array}{cc}
        -\frac{n}{2}\quad \frac{n}{2}+1+\frac{3m}{2}\quad \frac{1}{2}+i\,u\quad \frac{1}{2}-i\,u \\
        1+\frac{m}{2} \quad 1+\frac{m}{2}\quad 1+\frac{m}{2}
      \end{array}
    \right| 1\right).
\end{equation}
obeys the Baxter equation
\begin{eqnarray}
  && 
  \left(u-i\frac{m+1}{2} \right)^3 F_{n,m}(u-i)
  +\left(u+i\frac{m+1}{2} \right)^3 F_{n,m}(u+i) =
  t_3\left(u\right) F_{n,m}(u),
  \nonumber \\
  &&
  t_3\left(u\right) = 2 u^3-\left(n^2 -n+ 3 (m+1) n+\frac{3}{2} (m+1)^2\right)\,u .
\end{eqnarray}
In the formula for $Q_6$, we defined
\begin{eqnarray}
  \label{auxiliary-funct}
  && f_l\left(u\right)=-\frac{P_{l}\left(u-\frac{i}{2}\right)}{P_{l+1}\left(u\right)}
  \qquad \qquad
  \tilde{f}_l\left(u\right)=-\frac{P_{l}\left(u+\frac{i}{2}\right)}{P_{l+1}\left(u\right)},
\end{eqnarray}
and 
\begin{eqnarray}
  \label{auxiliary-funct2}
  P_{l}(u) & = & \sum_{k=0}^{m-1-l} \left(-1\right)^k \binom{m-1-l}{k} 
  \prod_{j=1}^{k}   \left(u -i \frac{2j+l}{2}\right)^3\times \\
  &&\times 
  \prod_{j=1}^{m-1-k-l} \left( u +i \frac{2j+l}{2}\right)^3
  F_{n,m}\left(u+i\frac{m-1-l-2k}{2}\right).
  \nonumber 
\end{eqnarray}

\section{Expansion of various hypergeometric functions}
\label{app:exp}

Let 
\begin{equation}
      L_{n,m}(u) = \log F_{n,m}(u).
\end{equation}
We can easily obtain closed expressions for the specialized derivatives
\begin{equation}
L_{n,m}^{(k)}\left(\frac{m+1}{2}\,i\right).
\end{equation}
In particular, the first two derivatives for $m=2,3,4$ are
\begin{eqnarray}
L'_{n,2}(3i/2) &=& -\frac{i \left(4 (n+4) S_1\left(\frac{n}{2}+1\right)-5 n-16\right)}{2 (n+4)}, \\
L''_{n,2}(3i/2) &=& \frac{n (n+8)}{4 (n+4)^2},
\end{eqnarray}
\begin{eqnarray}
L'_{n,3}(2\,i) &=&  -i S_1\left(\frac{n}{2}+2\right)-i S_1\left(\frac{n}{2}+4\right)-\frac{1}{2} i (4 \log (2)-9),\\
L''_{n,3}(2\,i) &=& S_2\left(\frac{n}{2}+2\right)-S_2\left(\frac{n}{2}+4\right)+\frac{1}{12} \left(4 \pi ^2-37\right),
\end{eqnarray}
\begin{eqnarray}
L'_{n,4}(5i/2) &=&  -i S_1\left(\frac{n}{2}+2\right)-i S_1\left(\frac{n}{2}+4\right)+\frac{43 i}{12},\\
L''_{n,4}(5i/2) &=& \frac{n \left(25 n^3+700 n^2+6148 n+17472\right)}{144 \left(n^2+14 n+48\right)^2}.
\end{eqnarray}
We can now use the Baxter equation to shift the arguments and move them to $i/2$ for even $m$ of $i$ for odd $m$.
Expanding at large $n$, we find
\begin{eqnarray}
\bar{n} &=& \frac{1}{2} e^{\gamma_{E}}\,n, \\
L'_{n,2}(i/2) &=& \left(3 i-2 i \log \left(\bar{n}\right)\right)-\frac{8 i}{n}+\frac{\frac{62 i}{3}-8 i \log \left(\bar{n}\right)}{n^2}+\frac{64 i \log
   \left(\bar{n}\right)-112 i}{n^3}+\dots, \nonumber \\
L''_{n,2}(i/2) &=& -3+\frac{32 \log ^2\left(\bar{n}\right)-16 \log \left(\bar{n}\right)+4}{n^2}+\frac{-256 \log ^2\left(\bar{n}\right)+384 \log
   \left(\bar{n}\right)-96}{n^3}+\dots,\nonumber
\end{eqnarray}
\begin{eqnarray}
\bar{n} &=& e^{\gamma_{E}}\,n, \\
L'_{n,3}(i) &=& \left(\frac{9 i}{2}-2 i \log \left(\bar{n}\right)\right)-\frac{11 i}{n}+\frac{217 i}{6 n^2}-\frac{176 i}{n^3}+\dots, \nonumber \\
L''_{n,3}(i) &=& \left(\frac{\pi ^2}{3}-\frac{15}{4}\right)-\frac{2}{n^2}+\frac{22}{n^3}+\dots, \nonumber
\end{eqnarray}
\begin{eqnarray}
\bar{n} &=& \frac{1}{2} e^{\gamma_{E}}\,n, \\
L'_{n,4}(i/2) &=& \left(\frac{9 i}{2}-2 i \log \left(\bar{n}\right)\right)-\frac{14 i}{n}+\frac{\frac{446 i}{3}-64 i \log \left(\bar{n}\right)}{n^2}+\frac{896 i
   \log \left(\bar{n}\right)-2072 i}{n^3}+\dots, \nonumber \\
   L''_{n,4}(i/2) &=& -\frac{15}{4}+\frac{256 \log ^2\left(\bar{n}\right)-704 \log \left(\bar{n}\right)+536}{n^2}+\frac{-3584 \log ^2\left(\bar{n}\right)+13440 \log
   \left(\bar{n}\right)-12432}{n^3}+\dots. \nonumber
\end{eqnarray}

\end{document}